\documentclass[pra,aps,twocolumn,epsf]{revtex4}%
\usepackage{amsmath}
\usepackage{graphicx}
\usepackage{amsfonts}
\usepackage{amssymb}%
\begin{document}
\title{Factoring numbers with a single interferogram}
\author{Vincenzo Tamma$^{1,2}$}
\author{Heyi Zhang$^{1}$}
\author{Xuehua He$^{1}$}
\author{Augusto Garuccio$^{2}$}
\author{ Wolfgang P. Schleich$^{3}$}
\author{Yanhua Shih$^{1}$}
\affiliation{$^{1}$Department of Physics, University of Maryland, Baltimore County, Baltimore, Maryland 21250,  
$^{2}$Dipartimento Interateneo di Fisica, Universit\`{a} degli Studi di Bari, 70100 Bari, Italy, 
$^{3}$INFN – Sezione di Bari, Bari, Italy,
$^{4}$Institut f\"{u}r Quantenphysik, Universit\"{a}t Ulm, Albert-Einstein-Allee 11, D-89081 Ulm, Germany}

\begin{abstract}
\noindent
We construct an analogue computer based on light interference to encode the hyperbolic function $f(\zeta)\equiv 1/\zeta$ into a sequence of skewed curlicue functions. The resulting interferogram when scaled appropriately allows us to find the prime number decompositions of integers. We implement  this idea exploiting polychromatic optical interference in a multi-path interferometer and factor seven digit numbers. We give an estimate for the largest number that can be factored by this scheme.

\end{abstract}
\maketitle

To find the factors of a large integer number $N$ is a problem of exponential complexity. Indeed, the security of codes relies on this fact but is endangered by Shor's algorithm \cite{shor}, which employs entanglement between quantum systems \cite{schleich}. In the present paper we report the optical realization of a new algorithm for factoring numbers, which takes advantage of interference only.

A naive way of approaching the problem of factorization consists of dividing $N$ by integers $\ell$, starting from $\ell=3$ until $N/\ell$ is an integer. In the worst case this procedure requires $\sqrt{N}$ divisions before one would find a factor. On a digital computer division of large numbers is a rather costly process. However, in many physical phenomena division occurs in a rather natural way. For example, a wave of wavelength $\lambda$, propagating over a distance $L$, acquires a phase $\phi = 2 \pi L/\lambda$ and therefore probes the ratio $L/\lambda$. In the optical domain $\lambda$  is measured in nanometers ($nm$), that is $\lambda=\ell$ nm.  When we also express the path length $L$ in units of nm, that is $L=N$ nm, the phase $\phi = 2 \pi N/\ell$ is sensitive to the ratio $N/\ell$. For factors of $N$, $\phi$ is an integer multiple of $2\pi$. Otherwise $\phi$ is a rational multiple of $2\pi$.

In order to  enhance the signal associated with a factor relative to the ones corresponding to non-factors, we use interference of waves, which differ in their optical path length by an integer multiple. In this way we take advantage of constructive interference when $\ell$ is a factor of $N$, but destructive interference when $\ell$ is not a factor of $N$. The cancellation of terms is most effective when the individual optical paths increase in a nonlinear way. In this case, the intensity of the interfering waves is determined by the absolute value squared of a truncated exponential sum \cite{expsum}.  A polychromatic source of light, which contains several wavelengths $\lambda=\ell$ nm, allows us to test several trial factors simultaneously, taking advantage of the properties of truncated exponential sums \cite{continuousgauss} with continuous arguments.

Our method is motivated by recent work on factorization using truncated exponential sums \cite{merkel2}, which has been realized  in several experiments \cite{nmr1}. However, it differs from the past realizations in three important points:(i) the division of $N$  by the trial factors $\ell$ is not pre-calculated \cite{jones}, but it is performed by the experiment itself, (ii) all the trial factors are tested simultaneously in a single experiment,  and (iii) a scaling property inherent in the recorded interferogram of a single number allows us to obtain the factors of several numbers.

\begin{figure}[t]
   \begin{center}
   \begin{tabular}{c}
   \includegraphics[width=9cm]{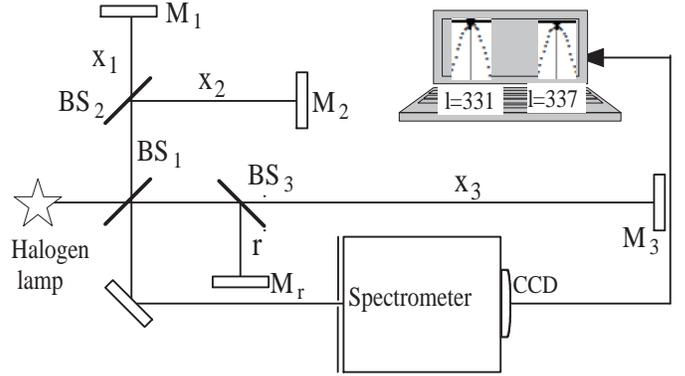}
   \end{tabular}
   \end{center}
\caption {Experimental setup for factoring numbers using classical interference in a generalized symmetric Michelson interferometer with $M+1$ paths. This analogue computer consists of a polychromatic source (halogen lamp),  $M$ balanced beam splitters, $M+1$ mirrors, and a spectrometer connected to a CCD camera. The $M$
interfering paths are varied with respect to the reference path of length $r$, defined by the reference mirror $M_{r}$ and indicated by thin vertical dashed lines, by displacing the $M$ mirrors. The length of the $m$-th arm reads $x_{m} = r + (m-1)^{2} x$, with the unit of displacement $x$ and $m=1,2,...,M$.  Here we depict the case $M=3$.  \label{setup}}
\end{figure}

The optical setup used to implement this idea is a symmetric Michelson interferometer with $M+1$ paths in free space shown in Fig. \ref{setup} for $M=3$.  In order to calibrate the interferometer the mirrors are arranged such that the path lengths are identical to the length $r$ of the reference path. Next we displace these mirrors in a nonlinear way as to obtain the final path length $x_{m}\equiv r + (m-1)^2 x$
of the $m$-th arm. Here $x$ denotes the unit of displacement. The preparation of the interferometer is completed after we have blocked
the reference mirror.

The intensity in the exit port of the interferometer is given by the interference of the waves in the remaining $M$ arms. Since we deal with balanced beam splitters all waves have the same amplitude. However, due to the different arm lengths $x_m$ of the interferometer the phases $\phi_m= 2\pi x_{m}/\lambda=2\pi r/\lambda + 2\pi(m-1)^2 x/\lambda$
give rise to the intensity
\begin{eqnarray}\label{interferogram}
I(\lambda;x) = |s(\frac{x}{\lambda})|^2,
\end{eqnarray}
which  we have expressed by the curlicue function \cite{berry}
\begin{eqnarray}\label{generalcgs}
s(\xi) \equiv \frac{1}{M}\sum_{m=1}^{M} \exp\left[
2\pi  i (m-1)^2 \xi\right].
\end{eqnarray}
Here we have normalized the output intensity with respect to the source intensity. Moreover, the reference phase $\phi_r\equiv 2\pi r/\lambda$ which is independent of $m$ has dropped out due to the fact that the intensity in Eq. (\ref{interferogram}) involves the  absolute value squared of the sum $s$.

From Eq. (\ref{generalcgs}) we note that $|s|^2$ has a dominant maximum at $\xi=0$ with $s(0)=1$ and decaying oscillations on the sides. Moreover, we recognize  the periodicity property $s(\xi + 1)=s(\xi)$. Therefore, it suffices to consider $s=s(\xi)$ in the interval $-1/2 \leq \xi \leq 1/2$. In addition $|s(\xi)|^2$ is symmetric with respect to $\xi=0$.

We now consider the dependence of the intensity $I$ given by Eq. (\ref{interferogram}) on the wavelength $\lambda$ for a fixed unit $x$ of displacement. The argument $\xi$ of $s$ is the ratio $x/\lambda=k(\lambda;x)+\tau(\lambda;x)$,
which we represent by the sum of the integer $k$ and the correction term  $\tau$ with $-1/2 \leq \tau \leq 1/2$. For a fixed value of $x$, $k$ and $\tau$ depend on $\lambda$. It is this dependence of $I=I(\lambda;x)$ on $\lambda$ which contains the information about the division by $\lambda$. Indeed, whenever $\lambda$ is such that $s=1$ we find $\tau=0$ and hence $x/\lambda=k$,
which implies that $\lambda$ is a factor of $x$.

Since $x$ and $\lambda$ only enter into the intensity as a ratio we find immediately the scaling law
\begin{eqnarray}
I(\lambda;x)=I(N \frac{\lambda}{x};N),
\end{eqnarray}
which suggests that we can find the factors of $N$ by rescaling the interferogram as a function of the dimensionless variable
\begin{eqnarray}\label{scaling}
\xi_N\equiv N \frac{\lambda}{x}.
\end{eqnarray}

Hence, we can exploit the wavelength $\lambda$ as well as the unit of displacement $x$ to scan the possible trial factors in the interval $1\leq \xi_N \leq \sqrt{N}$. In principle, by choosing a suitable value of $x$ and an appropriate interval of wavelengths we can factor any large value of $N$.

It is illuminating to compare our method to factor numbers to Shor's algorithm \cite{shor}. Both implement a function. We use a physical system, that is, a classical analogue computer to obtain the continuous function $f(\zeta)\equiv1/\zeta$. Shor's method takes advantage of the exponentially large Hilbert space of a quantum system to encode the discrete function $g(i)\equiv a^i mod N$ whose period shares a common factor with $N$. In contrast, we exploit the very change of the periodicity induced by the function $1/\zeta$ in the interferogram consisting of a sequence of skewed curlicue functions.

We now turn to the experimental implementation of our technique based on a symmetric Michelson interferometer with $M=3$ beam splitters and four mirrors.  Each mirror is mounted on a single axis translation stage which consists of a $5$ mm manual travel stage, a $50$ mm step motor, and $20$ $\mu$m piezoelectric and feedback control stage with a resolution of $10$  nm.

The polychromatic source of the interferometer is a halogen lamp with a bandwidth ranging from $400$ nm to $800$ nm. The interference pattern  at the output port is measured by a spectrometer \cite{note} connected to a $2048$ pixels Charge-Coupled Device (CCD) array of resolution $0.005-0.006$  nm,  as a continuous function of the wavelengths $\lambda$ associated with the polychromatic source. The spectrometer and the CCD can cover a range in wavelengths of the order of the light source. An accuracy of the order of $0.006$ nm in a single pixel requires the use of a CCD bandwidth of $13$ nm.

The calibration of the interferometer with a suitable accuracy is one of the challenging tasks in this experiment. We first determine when all path lengths $x_m$ are equal to $r$, by measuring the polychromatic two-path interference between the $m$-th beam and the reference beam, for each $m=1,2,3$, with the mirror $M_{r}$ tilted by a small angle with respect to all the other mirrors.
Then we block the mirror $M_r$ and obtain non-linear interfering path lengths by translating each mirror $M_{m}$ with the piezoelectric translators combined to the step motors. Our experiment uses the displacement unit $x=523426.8$  nm.

\begin{figure*}[t]
   \begin{center}
   \begin{tabular}{c}
   \includegraphics[width=14cm]{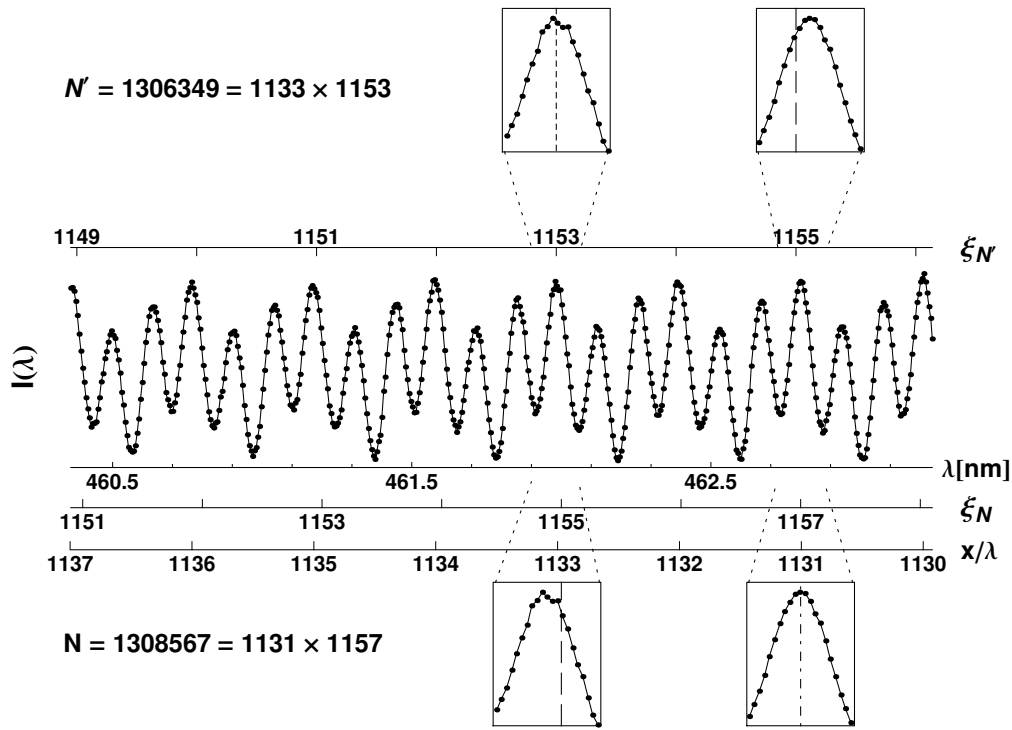}
   \end{tabular}
   \end{center}
\caption {Experimental interferogram $I=I(\lambda;x)$, obtained by the multi-path Michelson interferometer of Fig. 1, for a unit of displacement $x=523426.8$ nm, in the wavelength range $460.36$ nm$\leq \lambda \leq 463.24$ nm (center) and factorization of the two numbers $N=1308567=1131\times 1157$ (bottom) and $N'=1306349=1133\times1153$ (top), by rescaling the wavelength axis according to Eq. (\ref{scaling}). The dots represent the measured values and the curve is obtained by joining the experimental points.   The insets magnify the behavior of the interferogram in the neighborhood of the three trial factors $1153,1155$, and $1157$ whose locations are indicated by dotted, dashed and dashed-dotted lines, respectively. Only when the dominant maximum in the interferogram is located at an integer we have a factor of $N$ or $N'$. Every dominant maximum corresponds to a factor $x/\lambda$ as indicated by the horizontal axes at the bottom.
\label{experimentalresults}}
\end{figure*}

In the center of Fig. \ref{experimentalresults},
we show the resulting interferogram as a function of the wavelength $\lambda$ in the interval
$[460.36$ nm, $463.24$ nm$]$. Next we use this pattern  to factor two distinct numbers. We start with  $N=1308567=1131\times 1157$ and  present towards the bottom of Fig. \ref{experimentalresults} an axis with the variable $\xi_{N}$ rescaled according to Eq. (\ref{scaling}). We clearly identify the factor $1157$ by the maximum being located \cite{comment} at an integer, as shown by the inset. Moreover, we use the same interferogram to factor the number $N'=1306349=1133\times1153$. For this purpose we show  on the top  the rescaled variable $\xi_{N'}$. Again we can identify the factor $1153$ by the maximum being located at an integer.

What is the range of wavelengths necessary to factor a given number N? The answer to this question emerges from the condition that the spectrum must cover all test factors ranging from $1$ to $\sqrt{N}$, that is $1<\xi_N<\sqrt{N}$ together with the scaling transformation Eq. (\ref{scaling}) which translates into the wavelength requirement $1/N<\lambda/x<1/\sqrt{N}$.
Hence, for a given wavelength range $\lambda_{min}\leq \lambda \leq \lambda_{max}$ we can factor  numbers $N$ from the interval
\begin{eqnarray}\label{rangeN}
N_{min}\equiv(\frac{x}{\lambda_{max}})^2\leq N\leq\frac{x}{\lambda_{min}}\equiv N_{max},
\end{eqnarray}

We emphasize that this inequality also puts a restriction on the displacement $x$. Indeed, in order to have $N_{min}\leq N_{max}$ we need to satisfy the condition $(x/\lambda_{max})^2\leq x/\lambda_{min}$,
which implies
\begin{eqnarray}\label{condx}
x\leq\frac{\lambda_{max}^2}{\lambda_{min}}.
\end{eqnarray}

According to Eq. (\ref{rangeN}) the largest possible number that can be factored is given by the maximum value of $x$ which translates with the help of Eq. (\ref{condx}) into
\begin{eqnarray}\label{Nmax}
N_{max}=(\frac{\lambda_{max}}{\lambda_{min}})^2\equiv\beta^2.
\end{eqnarray}
With the wavelength domain of the halogen lamp we find from Eq. (\ref{Nmax}) the value $N_{max}=4$.

This conservative estimate originates from the constraint that we have to cover a wavelength domain large enough to include all test factors up to $\sqrt{N}$. Nonetheless we can still find the factors of some numbers as demonstrated by Fig. $2$. Indeed, each dominant peak in the interferogram corresponds to a factor $p$ of a number $N=p\cdot q$. Since the scaling law Eq. (\ref{scaling}) predicts $\xi_N=N \lambda/x=p\cdot q \lambda/x$ that is $q=x/\lambda$ the spectral range $\lambda_{min}\leq\lambda\leq\lambda_{max}$ allows us to determine the factors
\begin{eqnarray}\label{numberinterval}
q_{min} \equiv \frac{x}{\lambda_{max}}< q < \frac{x}{\lambda_{min}} \equiv q_{max}.
\end{eqnarray}
In order to illustrate this feature we have included in Fig. $2$ a horizontal axis on which we mark the possible factors $q$ given by the maxima of the curlicue function.

 One possibility of circumventing the bandwidth limitation given by Eq. (\ref{Nmax}) consists of repeating the experiment for different values of the unit $x$ of displacement while maintaining the bandwidth expressed by the dimensionless parameter $\beta$. For a given integer N to be factored we choose $n$ suitable values of $x=x_i$, with $i=0,1,…,n-1$ in order to cover subsequent intervals $[\xi_{N}^{(i)},\xi_{N}^{(i+1)}]$ of test factors with $\xi_{N}^{(i)}\equiv N \lambda_{min} /x_i$ such that $[\xi_{N}^{(0)},\xi_{N}^{(n)}]=[1,\sqrt{N}]$. We achieve this goal provided $x_{i+1} = x_i/\beta$, with $x_{0} \equiv  N \lambda_{min}$, and $n\lesssim log_{\beta} \sqrt{N}$ number of interferograms.

Similarly, we can factor any number in the interval $N_{min} < N < N_{max}$ by using
$n\lesssim log_{\gamma} \sqrt {N_{max}}$ interferograms provided
$\gamma\equiv N_{min}/N_{max} \beta>1$.
Such interferograms are obtained for values $x_i$ such that $x_{i+1}= x_i/\gamma<x_i$ with
\begin{eqnarray}\label{x02}
x_{0} \equiv  N_{max} \lambda_{min}.
\end{eqnarray}

For example, for factoring all the integers N such that $N_{max}= 10 N_{min}$ it would be enough to exploit a wavelength bandwidth  $\beta=20$ in order to achieve the value $\gamma=2$.

In conclusion, we have outlined and verified by an experiment a new approach towards factoring numbers with the help of a multi-path Michelson interferometer where the individual optical path lengths increase quadratically leading to Gauss sums. However, we could have easily implemented any other polynomial increase where an exponential sum determines the output intensity of the interferometer. We have factored different numbers  with up to seven digits from a single recorded interferogram exploiting a remarkable scaling property.

Our experiment relied on only $M=3$ interfering paths. However, for larger numbers to be factored it can be useful to increase either the number of paths $M$ or the polynomial order of the interferograms in order to obtain sharper peaks and better accuracy in the value of the maxima. A  standard best fitting procedure \cite{comment} would allow us, in principle, to achieve the necessary accuracy even exploiting small values of $M$.  Nevertheless it is possible with $M=2$ paths to resolve consecutive peaks according to the Rayleigh criterion no matter how large the numbers to be factored.

It is amusing to consider the resources necessary to factor a $200$-digit number which is considered really hard on a classical computer. From  Eq. (\ref{x02}) we find for $\lambda_{min}\sim 100$ nm the estimate $x\sim 10^{53}$ m which is much larger than  the size of the universe \cite{univ} of $10^{27}$ m. Hence, such large numbers are not factorable with our method in the present form.

Obviously our analogue factorization algorithm does not provide an exponential speedup. Nevertheless it is very different from other classical algorithms which run on a digital computer. Indeed, in our case it is a physical phenomenon, that is interference, which we exploit to compute the factors of a given number. Although it is not possible to fully compare a digital with an analogue method we believe that our procedure paves the way for the use of physics for solving problems in number theory and we let ``nature'' solve complex problems for us.

The authors thank  J. Franson, K. McCann, A. Pittenger, T. Pittman, M. H. Rubin, H. Winsor and  T. Worchesky  for
useful suggestions and discussions. This research was
partially supported by AFOSR and ARO. 
V.T. acknowledges the joint Ph.D program at University of Maryland, Baltimore County and University of Bari for partial support.



\begin{thebibliography}{}
\bibitem{note} We calibrate the spectrometer by using light sources with well known emission spectra such as mercury lines and sodium double D lines. The accuracy in the calibration of the spectrometer is high enough within the allowed experimental error.
\bibitem{comment} In our experiment there was no ambiguity in determining  whether a dominant maximum is indeed located at an integer. Moreover, there was no need to normalize the output spectrum with respect to the source since the black body spectrum of the source is pretty constant in the measured bandwidth. However, for larger numbers we suggest the following strategy: (i) we first normalize the output pattern of the interferometer with respect to the intensity distribution of the source,  (ii) then perform a best fit of the experimental pattern, and (iii) read out all the value of the maxima by using a standard peak finding procedure and check for a each number $N$ to be factored if they  correspond within the experimental accuracy in the wavelengths to an integer value of $\xi_N$. Since some non-factors may become indistinguishable from the values of the maxima in the interferogram  we have to check  if they actually divide $N$.
\end{thebibliography}
\end{document}